\documentclass[amssymb,preprintnumbers,preprint, eqsecnum,amsmath,showpacs, nofootinbib]{revtex4}
\usepackage{graphicx,bm}
\newcommand{\bal}{\begin{equation}\begin{aligned}}
\newcommand{\eal}{\end{aligned}\end{equation}}

\newcommand{\ri}{r_\infty}
\begin{document}
\vspace*{5cm}

\bigskip
\title{ \Large
Thermodynamics of Squashed Kaluza-Klein Black Holes and Black 
Strings \\
-- A Comparison of Reference Backgrounds --
\vspace{1cm}
}

\author{\Large  
	Yasunari Kurita\footnote{E-mail:kurita@sci.osaka-cu.ac.jp} 
	}
  	
\address{ \bigskip
	Osaka City University Advanced Mathematical Institute, 
  	Osaka 558-8585, Japan
% \vspace{2cm}
  }
   \author{\Large
        Hideki Ishihara\footnote{E-mail:ishihara@sci.osaka-cu.ac.jp} 
        }
  \address{
\bigskip 
        Department of Mathematics and Physics, Osaka City University,
  	Osaka 558-8585, Japan
  }

\begin{abstract}
We investigate thermodynamics constructed on different
background reference spacetimes for 
squashed Kaluza-Klein (SqKK) black hole
and electrically charged black string in five-dimensional Einstein-Maxwell system. 
Two spacetimes are possible to be reference spacetimes 
giving finite gravitational classical actions: 
one is four-dimensional Minkowski times a circle and the other is the KK monopole. 
The boundary of the SqKK black hole can not be matched 
perfectly to that of the former reference spacetime because of the 
difference in topology. 
However, the resultant classical action coincides with 
that calculated by the counterterm subtraction scheme. 
The boundary of the KK monopole has the same topology with that of the 
SqKK black hole and can be matched to the boundary 
of the black hole perfectly. 
The resultant action takes different value from 
the result given by using the former reference spacetime. 
After a brief review of thermodynamic quantities of the black hole solutions, 
we calculate thermodynamic potentials relevant for several thermodynamic 
environments.
The most stable state is different for each environment: 
For example, the KK monopole is the most stable state in 
isothermal environment with fixed gravitational tension. 
On the other hand, when the size of the extra-dimension is fixed,
the Minkowski times a circle is the most stable. 
It is shown that these two spacetimes can be reference spacetimes 
of the five-dimensional black string. 
\end{abstract}

\date{\today}

\preprint{OCU-PHYS 289}
\preprint{AP-GR 54}

\maketitle

%==============================================================
\section{Introduction}
\label{sec:intro}
%============================================================== 

The gravitational path integral in the saddle-point approximation has a long history in
the context of black hole thermodynamics~\cite{Gibbons:1976ue}. 
In general, there is a problem that must be faced with this approach, that is, the 
classical action diverges. 
Thus, in order to obtain finite action, an appropriate regularization is needed.
There are two well-known regularization schemes: \lq\lq background subtraction"
which is a traditional approach and the \lq\lq counterterm subtraction".

In the background subtraction scheme, we consider the values of the action relative to 
a background reference spacetime,
where the choice of the reference spacetime is ambiguous in some cases. 
An example is the Euclidean Taub-bolt solution; 
both Minkowski spacetime \cite{Gibbons:1979nf} and the self-dual Taub-NUT~\cite{Hunter:1998qe}
are possible to be reference spacetimes. 
In the case of Minkowski spacetime, the asymptotic boundary can not be matched perfectly to that of 
the Taub-bolt because of the difference in topology, 
whereas the boundary of the self-dual Taub-NUT solution has the same 
topology and can be matched perfectly.
The resulting finite actions evaluated on these two reference spacetimes are different.

Motivated by the AdS/CFT correspondence~\cite{Malda,WittenAdS/CFT,GKP}, 
the counterterm subtraction method was developed by Balasubramanian and Kraus~\cite{Balasubramanian:1999re},
who proposed adding a counterterm to gravitational action at asymptotic boundary. 
One remarkable thing is that this method does not require to choose any reference spacetime and there is no
ambiguity mentioned above.
The counterterm method was found originally for asymptotically AdS 
or asymptotically locally AdS spacetime 
\cite{Balasubramanian:1999re}, 
and then extended for a class of $(d+1)$-dimensional asymptotically 
flat spacetime with boundary $S^n\times R^{d-n}$
\cite{Kraus:1999di,Lau:1999dp,Mann:1999pc}
and five-dimensional asymptotically locally flat spacetime 
with a fibered boundary topology $R^2 \hookrightarrow S^2$ in \cite{Mann:2005cx}.
This progress makes it possible to investigate gravitational mass of 
asymptotically locally flat spacetime like the Kaluza-Klein (KK) 
monopole \cite{Gross-Perry,Sorkin} as was done 
in \cite{Mann:2005cx}.

By applying the counterterm method to Taub-bolt-AdS and taking zero cosmological constant limit 
after the evaluation of the gravitational action, in \cite{Emparan:1999pm},
Emparan, Johnson and Myers compared the action with that computed by the background subtraction method. 
They showed that the resultant action of Taub-bolt instanton by counterterm method 
agrees precisely with the one by background subtraction method with the different 
boundary topology given in~\cite{Gibbons:1979nf}.
Thus, the generating functional in the CFT side seems to prefer the 
gravitational action evaluated using the spacetime boundary with 
different topology. 
This might imply that, in the background subtraction scheme, 
topology of the asymptotic boundary is not so important.

Recently, for the purpose of searching the end state 
of the Gregory-Laflamme instability\cite{Gregory:1993vy,Gregory:1994bj}, 
much effort has been devoted to 
the investigation of the so-called caged Kaluza-Klein(KK) black holes, 
which are defined to be solutions with a spherical event horizon that asymptote to 
Minkowski space times a circle 
(See {\it e.g.} \cite{Kol:2004ww,Harmark:2007md} 
and the references therein.).
The thermodynamic first law of the caged KK black holes was obtained in
\cite{Townsend:2001rg,Kol:2003if,Harmark:2003eg}, showing that 
the caged KK black holes can be characterized by mass and gravitational 
tension.

In the five-dimensional Einstein-Maxwell theory, 
there are analytic solutions representing 
electrically charged black holes with squashed horizons 
\cite{Ishihara-Matsuno} as a generalization of the solutions 
given in \cite{Dobiasch:1981vh} and \cite{Gibbons:1985ac}. 
Intriguingly, the spacetime far from the black hole is 
locally a product of the four-dimensional Minkowski spacetime and a 
circle\footnote{The circle is fibered over the four-dimensional spacetime
with non-trivial twisting.}.
In this sense, the black hole resides in KK spacetime and 
is worth to be named a squashed Kaluza-Klein (SqKK) black hole. 
The dynamical stability of the SqKK black hole was investigated in \cite{Kimura07}. 
The thermodynamic properties of the SqKK black hole was investigated 
by use of the counterterm method in~\cite{Cai:2006td} 
and by the background subtraction scheme in \cite{Kurita:2007hu}.
The thermodynamics of the rotating uncharged SqKK black hole was 
also investigated in \cite{Wang:2006nw}.

When one try to evaluate the classical action of the SqKK 
black hole by the background subtraction scheme, 
there is an ambiguity in the choice of reference spacetime. 
As far as we are aware, there are two different reference spacetimes; 
one is Minkowski spacetime times $S^1$ and the other is the KK monopole.
In this paper, we call the former reference spacetime the flat 
background, in short. 
The asymptotic boundary of the flat background is not the same as 
that of the SqKK black hole in topology, 
but the boundary of the KK monopole has the same topology and can be 
matched to that of the SqKK black hole perfectly.
Therefore, in the counterterm method, 
the counterterm relevant for the KK monopole is applicable also to the 
SqKK black hole spacetime.

In this paper, we compare these two reference spacetimes from 
the viewpoint of thermodynamics. 
In the background subtraction scheme, free energy 
can be interpreted as a relative value to the reference background 
which one chooses. 
We calculate free energy, or equivalently, thermodynamic potentials 
in various thermodynamic environments 
characterized by thermodynamic variables obtained in~\cite{Kurita:2007hu}. 
It will be shown that the most stable state among these solutions depends on 
the choice of thermodynamic environment. 
For example, KK monopole can be the most stable state 
in isothermal environments with fixed gravitational tension, 
but the flat background is the most stable 
in isothermal environments with fixed size of the extra-dimension at 
spatial infinity.

We will also show that these two reference spacetimes can be reference spacetimes of
five-dimensional electrically charged black string.
We calculate its free energy and thermodynamic potentials in some thermodynamic environments.

The organization of the present paper is as follows.
In the next section, we review the SqKK black hole and 
give the metrics of the two reference spacetimes. 
In section \ref{sec:actions}, calculations of 
the classical actions are given in detail by two methods:
the background subtraction method with respect to the two reference spacetimes 
and the counterterm method. 
In section \ref{sec:themodynamical-relation}, we give a brief review of 
the thermodynamic quantities of the SqKK black hole and expressions 
of the first law obtained in \cite{Kurita:2007hu}.
In section \ref{section:potentials}, thermodynamic potentials relevant for thermodynamic environments 
are calculated and the most stable state in each environment is clarified.
In section \ref{section:BS}, using these two reference spacetimes, 
we apply the background subtraction method to the five-dimensional charged black string. 
In section \ref{section:summary}, we summarize the results and discuss a remaining problem.

%%%%%%%%%%%%%%%%%%%%%%%%%%%%%%%%%%%%%%%%%%%%%%%%%%%%%%%%%%%%%%%%%%%%%%%%%%%%%%%%%%%%%%%%%%%%%%%%%%%%%%%
\section{The solution}
\label{sec:solution}
%%%%%%%%%%%%%%%%%%%%%%%%%%%%%%%%%%%%%%%%%%%%%%%%%%%%%%%%%%%%%%%%%%%%%%%%%%%%%%%%%%%%%%%%%%%%%%%%%%%%%%%

Let us review the SqKK black hole~\cite{Ishihara-Matsuno}, 
which is a solution of the five-dimensional Einstein-Maxwell theory. 
The action is given by 
 %----------------------------------------------------
 \begin{eqnarray}
 I = \frac{1}{16\pi G} \int_{\cal{M}} d^5 x \sqrt{-g} 
         \left[ R - F^{\mu\nu} F_{\mu\nu} \right]+\frac{1}{8\pi G} \int_{\partial \cal{M}} 
 K\sqrt{-h}d^4x, 
 \end{eqnarray}
 %---------------------------------------------------
where $G$ is the five-dimensional Newton constant, 
$g_{\mu\nu}$ is the metric, $R$ is scalar curvature, 
$F_{\mu\nu} = \partial_\mu A_\nu - \partial_\nu A_\mu $ is 
the field strength of the five-dimensional $U(1)$ gauge field $A_\mu$. 
The second term is so-called Gibbons-Hawking term, 
in which $h$ is determinant of the induced metric and 
$K$ is the trace of the extrinsic curvature of the boundary 
$\partial \cal{M}$, respectively. 
It is required so that 
upon variation with metric fixed at the boundary, 
the action yields the Einstein equations \cite{Gibbons:1976ue}. 
The metric of the SqKK black hole considered in this paper is given by 
% %-------------------------------------------------
\begin{eqnarray}
 ds^2 &=& - V(\rho) d\tau^2 + \frac{B (\rho)}{V(\rho)} d\rho^2 
 + \rho^2 B (\rho) d\Omega^2  
 + \frac{r_{\infty}^2}{4 B (\rho)} \left(d\psi + \cos \theta d\phi \right)^2 ,
\label{eq:metric-Trho}
\end{eqnarray}
%-------------------------------------------------
where $d\Omega^2 = d\theta^2 + \sin^2 \theta d\phi^2$ is the metric of the unit two-dimensional sphere and 
%------------------------------------------------
\begin{eqnarray}
V(\rho)=  \left( 1- \frac{\rho_{+}}{\rho}\right)\left( 1-\frac{\rho_{-}}{\rho} \right) , \ \ 
B(\rho) =   1+ \frac{\rho_0}{\rho}, \ \ 
r_{\infty} = 2\sqrt{(\rho_+ +\rho_0 )(\rho_- + \rho_0 )}. 
\end{eqnarray}
%-------------------------------------------------
Here, the coordinate ranges are 
$0 \leq \theta <2\pi ,\   0 \leq \phi < 2\pi ,\  0 \leq \psi < 4\pi $. 
The gauge potential is given by
%----------------------------------------
\begin{eqnarray}
A = \mp   \frac{\sqrt{3}}{2}  \frac{\sqrt{{\rho_+}{\rho_-}}}{\rho} d\tau.
\label{eq:-Maxwell-gauge-field}
 \end{eqnarray}
%--------------------------------------------
When $\rho_0>0$,  we consider parameter region $\rho_+ \geq \rho_- \geq 
0$, which does not lose any generality. 
We note that even when $\rho_0$ is negative, the metric (\ref{eq:metric-Trho}) describes 
a black hole geometry if the parameters satisfy $-\rho_0=|\rho_0|< 
\rho_-\leq \rho_+$~\cite{inner}.

It is easy to see the apparent singularity at $\rho_{+}$ corresponds to 
the outer horizon of the SqKK black hole.
The inner horizon at $\rho_{-}$ is analogous to that of the 
Reissner-Nordstr\"om black holes. 
It should be noted that the shape of the horizon is 
a squashed $S^3$ as was discussed in~\cite{Ishihara-Matsuno}.
The spatial infinity corresponds to a limit $\rho \to \infty$. 
In the limit, the metric asymptotes to
%------------------------------------------------
\begin{eqnarray} 
ds^2=-d\tau^2+d\rho^2+\rho^2d\Omega^2+\frac{r_{\infty}^2}{4}
(d\psi + \cos\theta d\phi)^2, 
\label{eq:metric-Lf}
\end{eqnarray} 
%----------------------------------------------------
which is a metric of a \lq\lq twisted'' $S^1$ fiber bundle over 
four-dimensional Minkowski spacetime. 
From the metric (\ref{eq:metric-Lf}), 
it is seen that the $S^1$ circle parametrized by a coordinate $\psi$ 
has finite size even at the spatial infinity. 
The non-trivial twisting of the $S^1$ circle fibered over the $S^2$ base space 
leads a 4-dimensional U(1) gauge field by KK reduction. 
Actually, in no horizon limit $\rho_{\pm} \to 0$ with $\rho_0 >0$, 
the SqKK black hole spacetime becomes the KK monopole spacetime 
\cite{Gross-Perry,Sorkin}. 
The metric of the KK monopole is
%----------------------------------------------------------------------
\begin{eqnarray}
ds^2 &=& - d\tau^2 + \left( 1+ \frac{\rho_0}{\rho} \right) d\rho^2 
 + \rho  \left( \rho+ {\rho_0} \right)  d\Omega^2  
 + \frac{\rho_0^2 {\rho}}{ ( \rho+ {\rho_0}) } 
\left(d\psi + \cos \theta d\phi \right)^2,
\label{eq:metric-GPS-Trho}
\end{eqnarray}
%------------------------------------------------------------------------------
which is a vacuum solution of the five-dimensional Einstein gravity.
In the limit $\ri \to \infty$, 
the KK monopole becomes five-dimensional Minkowski spacetime and 
the SqKK black hole reduces to the five-dimensional Reissner-Nordstr\"om 
black hole (For details, see \cite{Ishihara-Matsuno}).
Thus, this limit is the five-dimensional spherically symmetric limit, i.e., 
the spacetime admits SO(4) isometry group.

The metric (\ref{eq:metric-Lf}) has a curvature singularity at $\rho=0$ 
because of the non-trivial twisting. 
Therefore, the action and Hamiltonian\footnote{
In this paper, we consider the definition of the Hamiltonian given in 
\cite{Hawking:1995fd}.} diverge and 
the spacetime is not useful as a reference spacetime to regularize the 
action and Hamiltonian of the SqKK black hole. 
If the $S^1$ fiber is untwisted, the spacetime becomes a product $S^1$ 
bundle over four-dimensional Minkowski, 
with the metric 
%-------------------------------------------------------------------------------------
\begin{eqnarray}
ds^2 = -d\tau^2+d\rho^2+\rho^2d\Omega^2+\frac{r_{\infty}^2}{4}d\psi^2.
\label{eq:flatBGnonE}
\end{eqnarray} 
%---------------------------------------------------------------------------------
This is a completely flat solution of the five-dimensional vacuum Einstein equations.  
Thus, the flat spacetime is useful as a reference spacetime 
in evaluation of the action and the Hamiltonian. 
This is the flat background mentioned in section \ref{sec:intro}.

Given the metric (\ref{eq:metric-Trho}), we can calculate various physical quantities.
The surface gravity is calculated as
%------------------------------------------
\begin{eqnarray}
   \kappa_{+} = \frac{\rho_{+} -\rho_{-}}{2\rho_{+} \sqrt{{\rho_{+}}({\rho_{+} + \rho_0})}},
   \label{eq:surfacegravity}
\end{eqnarray}
%--------------------------------------
which gives the Hawking temperature of the SqKK black hole as 
$T= \kappa_+/2\pi$ \cite{Hawking:1974sw}.
The greybody factor of the SqKK black hole was investigated in \cite{Ishihara:2007ni,Chen:2007pu}.
We assume that the entropy of the SqKK black hole is given by 
the Bekestein-Hawking formula~\cite{Bekenstein:1973ur,Hawking:1974rv}
%-------------------------------------
\begin{eqnarray} 
S = \frac{A_+}{4G} = \frac{4\pi^2}{  G}  \rho_+  (\rho_+ + \rho_0) \sqrt{\rho_+(\rho_- + \rho_0 ) },
 \label{eq:entropy}
\end{eqnarray}
%-------------------------------------
where $A_+$ is the area of the outer horizon,
which is consistent with the Wald's entropy formula \cite{Wald:1993nt,Iyer:1994ys}. 
The electric charge and electrostatic potential of the SqKK black hole 
are also calculated as \cite{Ishihara-Matsuno}
%--------------------------------- 
\begin{eqnarray}
  Q & = & \frac{1}{8\pi G} \int_\infty  dS_{\mu\nu} F^{\mu\nu}
     = 
    \pm \frac{\sqrt{3} \pi}{G} r_\infty \sqrt{\rho_{+} \rho_{-} }, \quad 
    \label{eq:charge} \\
\Phi  &=&  A_{\tau}|_{\rho \to {\infty}} - A_{\tau}|_{\rho=\rho_+} 
= \pm \frac{\sqrt{3}}{2} \sqrt{\frac{\rho_-}{\rho_+} }. 
 \label{eq:potential} 
\end{eqnarray}
%---------------------------
The Komar mass is a meaningful mass of black holes which possess a timelike Killing vector.
Using the timelike Killing vector $\partial_{\tau}$, 
we can calculate the Komar mass of the SqKK black hole as 
%----------------------------------------------------------------------------------
\begin{eqnarray}
M_K &=& -\frac{3}{32 \pi G} \int dS_{\mu\nu} D^{\mu}\xi^{\nu} = 
\frac{3\pi}{4G } (\rho_+ + \rho_-)r_{\infty}, 
\label{eq:Komar-rho} 
\end{eqnarray} 
%-------------------------------------------------------------------------------------
where the integral is taken over the squashed three-dimensional sphere 
at the spatial infinity. 
The Smarr-type formula was shown generally in~\cite{Gauntlett:1998fz} as 
 %---------------------------------------------------------
 \begin{eqnarray}
 M_K -Q\Phi = \frac{3}{2} TS,
 \label{eq:Smarr}
 \end{eqnarray}
 %-------------------------------------------------------
which is sometimes called integrated expression for the first law.

The Abbott-Deser mass (AD mass)~\cite{Abbott:1981ff} is 
also a meaningful mass of the SqKK black hole as was discussed in \cite{Cai:2006td}.
It is a definition of mass for spacetimes with arbitrary asymptotic behaviour\footnote{
As for the definition of the AD mass for the SqKK black hole, see \cite{Cai:2006td}, for example.}.
Using the flat background\footnote{In the calculation of the AD mass in \cite{Cai:2006td}, 
the locally flat spacetime with the metric (\ref{eq:metric-Lf}) was chosen as the reference spacetime.
However, as mentioned before, the locally flat background is not suitable 
for a reference spacetime in the evaluation of the classical action and 
the Hamiltonian. 
The flat background, instead, 
is a suitable reference spacetime for all of them. 
For maintaining consistency in the
choice of reference spacetime, we consider the flat background in this paper.}, 
the AD mass is calculated as 
%------------------------------------------------------------------
\begin{eqnarray}
 M_{AD} = \frac{\pi}{2G} \left( 2\rho_+ + 2\rho_- +\rho_0 \right)r_{\infty}.
\label{eq:ADmass-BH-flat}
\end{eqnarray}
%----------------------------------------------------------------------- 
For asymptotically flat stationary spacetimes, 
the Komar mass equals the ADM mass as far as we are aware.
However, interestingly, for the SqKK black hole we consider here, which is not asymptotically flat, 
the AD mass has a different value from the Komar mass.

%%%%%%%%%%%%%%%%%%%%%%%%%%%%%%%%%%%%%%%%%%%%%%%%%%%%%%%%%%%%%%%%%%%%%%%%%%%%%%%%%%%%%%%%%%%%%%%%
\section{The classical action and free energy of the SqKK black hole}
\label{sec:actions}
%%%%%%%%%%%%%%%%%%%%%%%%%%%%%%%%%%%%%%%%%%%%%%%%%%%%%%%%%%%%%%%%%%%%%%%%%%%%%%%%%%%%%%%%%%%%%%%%

The Euclidean action is
%----------------------------------------------------
\begin{eqnarray}
I_E = \frac{1}{16\pi G} \int_{\cal{M}} \left( R_E-F_{\mu\nu}F^{\mu\nu}\right)\sqrt{g_E} d^5x  
+ \frac{1}{8\pi G} \int_{\partial \cal{M}} K\sqrt{h}d^4x,
\end{eqnarray} 
%---------------------------------------------------
where $g_E$ is the Euclidean metric and $R_E$ is the scalar curvature.
As mentioned earlier, in order to evaluate a finite classical action,
we need some regularization scheme\footnote{
Such regularization is almost always needed for non-compact spacetime. 
As far as we know, the only exceptional case is lower-dimensional braneworld 
black hole spacetime. 
For details, see Kudoh and Kurita \cite{KK}.}.

Firstly, let us consider the background subtraction method, 
which requires a background spacetime as a reference. 
We set the boundary of the SqKK black hole spacetime, $\partial {\cal{M}}$, 
at $\rho=\rho_B=const.$, 
and calculate $I_E({\cal{M}}, \partial {\cal{M}})$. 
Similarly, for the reference spacetime ${\cal{M}}_{R}$ with 
$\partial {\cal{M}}_{R}$, we obtain $I_E({\cal{M}}_R, \partial {\cal{M}}_R)$.
After the subtraction 
$I_E({\cal{M}}, \partial {\cal{M}})- I_E({\cal{M}}_R, \partial {\cal{M}}_R)$, 
we will take a limit $\rho_B \to \infty$.
In order to obtain the finite action, the reference spacetime is required to have   
the same proper length along the imaginary time, the same size of the $S^1$ fiber
and the same circumference radius of the $S^2$ base space as those of 
the SqKK black hole at the boundary. 
The period of the imaginary time, $\beta$, should be related to the inverse of the temperature as
%----------------------------------------------------------------------
\begin{eqnarray}
\beta = T^{-1} = \frac{4\pi\rho_{+} \sqrt{{\rho_{+}}({\rho_{+} + \rho_0})}}{\rho_{+} -\rho_{-}}.
\label{eq:beta}
\end{eqnarray}
%-------------------------------------------------------------------------
Thus, to match the period of imaginary time $\tau_E$ means 
that the two spacetimes have the same temperature.

Now, we consider the flat background with the following metric as a reference spacetime, 
%----------------------------------------------------------
\begin{eqnarray}
ds^2 = V(\rho_B)d\tau_E^2  +d{\rho}^2 + {\rho}^2 d\Omega^2 +\frac{{r}_{\infty R}^2}{4}d\psi^2,
\label{eq:flatBG}
\end{eqnarray}
%---------------------------------------------------------- 
where the parameter $r_{\infty R}$ should be determined by 
${r}_{\infty R}^2 = {r}_{\infty}^2 B(\rho_B)^{-1}$, and 
the imaginary time has a period $\tau_E \sim \tau_E + \beta$.
In order to match the circumference radius of the $S^2$ base space at the boundary, 
the boundary of the reference spacetime is set at the hypersurface 
$\rho= \rho_{BR} := \rho_B\sqrt{ B(\rho_B)}$. 
Therefore, the classical action of the SqKK black hole relative to the 
reference (\ref{eq:flatBG}) is calculated as
%-----------------------------------------------------------------------
\begin{eqnarray}
I_E = \beta \frac{\pi}{2G} (\rho_+ + \rho_0 ) r_{\infty},
\label{eq:Euclid-action}
\end{eqnarray}
%----------------------------------------------------------------------
which leads a free energy 
%--------------------------------------------------------------------------
\begin{eqnarray}
F= \frac{ I_E }{\beta}  
= \frac{\pi}{2G} (\rho_+ + \rho_0 ) r_{\infty}.
\label{eq:BHfree-energy-flat}
\end{eqnarray}
%----------------------------------------------------------------------------
This quantity can be interpreted as the free energy of the SqKK black 
hole relative to the flat background.
In the KK monopole limit $\rho_{\pm} \to 0$, 
the parameter $\rho_0$ becomes to $r_{\infty}/2$ and 
(\ref{eq:BHfree-energy-flat}) becomes 
%---------------------------------------------------------------------------------
\begin{eqnarray}
F \to \hat{F} = \frac{\pi}{4G}r_{\infty}^2,
\end{eqnarray}
%--------------------------------------------------------------------------------
which is the free energy of the KK monopole relative to 
the flat background\footnote{
In this paper, $\hat{A}$ means quantity $A$ of the KK monopole evaluated on 
the flat background.}.
This free energy equals one calculated by the counterterm method 
obtained in~\cite{Mann:2005cx}.

We should note that 
the $S^1$ fiber parametrized by $\psi$ in the flat background is not twisted. 
Therefore, the boundary of the flat background is topologically 
different from that of the SqKK black hole spacetime, 
where $S^1$ fiber is twisted. 
In spite of this difference in topology, we can obtain the finite result.

Since the boundary of the KK monopole background is the same as that of 
the SqKK black hole, the KK monopole would be a natural reference 
spacetime for the SqKK black hole. 
We consider the metric of the KK monopole 
%----------------------------------------------------------------------------------
\begin{eqnarray}
ds^2 = V(\rho_B)d\tau_E^2+B_R({\rho})d{\rho}^2+{\rho}^2 B_R({\rho})
d\Omega^2+\frac{\rho_{0 R}^2}{B_R({\rho})} \chi^2, 
\label{eq:metric-boundry-matched-GPS}
\end{eqnarray}
%--------------------------------------------------------------------------------------------------
where 
%---------------------------------------------------------------------------------------------------
\begin{eqnarray}
B_R({\rho}) = 1+\frac{\rho_{0R}}{{\rho}}.
\end{eqnarray}
%-------------------------------------------------------------------------------------
The boundary in the SqKK black hole spacetime $\rho=\rho_B$ is matched to 
the surface ${\rho}=\rho_{BR}$ in the KK monopole background. 
The parameters $\rho_{0R}$ and $\rho_{BR}$ should satisfy 
%-----------------------------------------------------------------------------------
\begin{eqnarray}
\rho_B^2 B(\rho_B) = \rho_{BR}^2 B_R(\rho_{BR}), 
\quad \frac{r_{\infty}^2}{4B(\rho_B)} = \frac{\rho_{0R}^2}{B_R(\rho_{BR})}.
\end{eqnarray}
%----------------------------------------------------------
Then, we have 
%-----------------------------------
\begin{eqnarray}
\rho_{0R} = \frac{r_{\infty} \rho_B }{2}\left[\rho_B^2 + \rho_0 \rho_B -\frac{1}{2}\rho_B r_{\infty} 
 \right]^{-1/2}, \quad 
\rho_{BR} =  \left[ \rho_B^2 + \rho_0 \rho_B -\frac{1}{2}\rho_B r_{\infty}  \right]^{1/2}.
\end{eqnarray}
%------------------------------------------------------------------------------------
With these parameters, 
the circumference radii of $S^2$ and the $\psi$ circle at the boundary are matched. 
Again, the imaginary time has the period $\beta$.
Then, the classical action of the SqKK black hole relative to the KK monopole 
background is calculated as 
%-----------------------------------
\begin{eqnarray}
\tilde{I}_E = \frac{\pi}{2G} \beta r_{\infty} 
 \left(\rho_+ + \rho_0 -  \frac{r_{\infty}}{2} \right),
 \label{eq:action-on-KKmonopole}
\end{eqnarray}
%-----------------------------------
which gives the free energy of the SqKK black hole evaluated on the KK 
monopole background as 
%-----------------------------------
\begin{eqnarray}
\tilde{F} = \frac{\tilde{I}_E}{\beta} = \frac{\pi}{2G} r_{\infty} 
 \left(\rho_+ + \rho_0 -  \frac{r_{\infty}}{2} \right).
\end{eqnarray}
%-----------------------------------
Of course, the above free energy vanishes in the KK monopole limit 
$\rho_{\pm}\to 0$ with $\rho_0>0$.

It is easily seen that the difference between $F$ and $\tilde{F}$ is a free energy of the KK monopole 
relative to the flat background, or equivalently,
%-----------------------------------------------------------------------------
\begin{eqnarray}
F = \tilde{F} +\hat{F},
\label{eq:F-relative}
\end{eqnarray}
%-----------------------------------------------------------------------------
as expected.
Therefore, it is checked that these free energies are well-defined as relative quantities between any two spacetimes.

Secondly, the action of the SqKK black hole can also be calculated by the counterterm method. 
The counterterm for five-dimensional asymptotically locally flat spacetime 
with a fibered boundary topology $R^2 \hookrightarrow S^2$ was proposed in \cite{Mann:1999pc} as 
%---------------------------------------------------------------------------------------------
\begin{eqnarray}
I_{ct} = \frac{1}{8\pi G} \int_{\partial M}  \sqrt{2{\mathcal{R}} }\sqrt{-h}\ d^4 x, 
\label{eq:counterterms}
\end{eqnarray}
%-----------------------------------------------------------------------------------------------------
where ${\mathcal{R}} $ is the Ricci scalar of the induced metric on the boundary. 
The result is 
%---------------------------------------------------------------------------------- ------------------
\begin{eqnarray}
I_E + I_{ct} = \beta \frac{\pi}{2G} (\rho_+ + \rho_0 ) r_{\infty}. 
\label{eq:action-ct} 
\end{eqnarray}
%-----------------------------------------------------------------------------------------------
Then, the free energy is obtained as 
%---------------------------------------------------------------------------------------
\begin{eqnarray}
F_{ct}=\frac{I_E+I_{ct}}{\beta}=\frac{\pi}{2G} (\rho_+ + \rho_0 ) r_{\infty},
\end{eqnarray}
%--------------------------------------------------------------------------------
which is equal to the free energy relative to the flat background 
(\ref{eq:BHfree-energy-flat})\footnote{The same finite action (\ref{eq:action-ct})
and free energy can be 
obtained by use of the counterterm proposed in \cite{Kraus:1999di}.}.

Furthermore, the AD mass of the SqKK black hole evaluated on the flat background 
(\ref{eq:ADmass-BH-flat})
is the same as the counterterm mass obtained in \cite{Cai:2006td}\footnote{
The counter-term mass can be calculated by use of the stress-energy tensor obtained 
in \cite{Astefanesei:2005ad}.}.
Therefore, for the SqKK black hole, 
the counterterm method is equivalent to the background subtraction method 
using the flat background as a reference spacetime.

%%%%%%%%%%%%%%%%%%%%%%%%%%%%%%%%%%%%%%%%%%%%%%%%%%%%%%%%%%%%%%%%%%%%%%%%%%%%%%%%%%%%%%%%%%%%%%%%
\section{Thermodynamical relation between Hamiltonian, Abbott-Deser mass and Komar mass}
\label{sec:themodynamical-relation}
%%%%%%%%%%%%%%%%%%%%%%%%%%%%%%%%%%%%%%%%%%%%%%%%%%%%%%%%%%%%%%%%%%%%%%%%%%%%%%%%%%%%%%%%%%%%%%%%

In this section, we review thermodynamic quantities of the SqKK black hole 
shown in \cite{Kurita:2007hu} 
and give more detailed description of their definitions and calculations. 
Firstly, we review the definitions of the Hamiltonian and the gravitational 
tension.

The Hamiltonian for Einstein-Maxwell system is~\cite{Hawking:1995fd}
%-------------------------------------------------------------------------------------------
\begin{eqnarray}
H := -\frac{1}{8\pi G} \int_{B_{\tau}}  \sqrt{\sigma} \left[
N_L k + u_i \left(\Theta^{ij} -\Theta h^{ij} \right) N_j + 
2 A_{\tau} F^{\tau i}u_i N_L \right],
\label{eq:def-H}
\end{eqnarray}
%----------------------------------------------------------------------------------------------
where $B_{\tau}$ is the boundary at infinity of a $\tau$ constant 
hypersurface $\Sigma_{\tau}$, $N_L$ is the lapse function, 
$\sqrt{\sigma}$ is the area element of $B_{\tau}$, 
$k$ is the trace of the extrinsic curvature of $B_{\tau}$
embedded in $\Sigma_{\tau}$, $u$ is the outward pointing unit normal to 
$B_{\tau}$, $\Theta_{ij}$ is the 
extrinsic curvature of $\Sigma_{\tau}$ embedded in the spacetime manifold. 
$\Theta$ is the trace of the extrinsic curvature.

The gravitational tension is a thermodynamic quantity of 
the SqKK black hole.
It was originally introduced for the black p-branes or black strings 
and contribute to their thermodynamical first law as 
an intensive quantity conjugate to the size of the compact dimension
\cite{Kol:2003if,Townsend:2001rg,Harmark:2003eg,Traschen:2001pb}. 
In~\cite{Harmark:2004ch}, the gravitational tension for a 
non-asymptotically flat spacetime relative to a reference spacetime was 
defined by using the Hamiltonian formalism to a foliation of the spacetime 
along asymptotically translationally-invariant spatial direction. 
The gravitational tension of the SqKK black hole along the $\psi$ 
direction is given by 
%---------------------------------------------------------------------------------------------
\begin{eqnarray}
\mathcal{T} := -\frac{1}{8\pi G \beta} \int_{B_{\psi}}  \sqrt{\bar{\sigma}} \left[
\bar{N}_L \bar{k} + \bar{u}_i \left(\bar{\Theta}^{ij} - \bar{\Theta} h^{ij} \right) \bar{N}_j
 + 2 A_{\psi} F^{\psi i}\bar{u}_i \bar{N}_L \right],
\label{eq:def-tension}
\end{eqnarray}
%----------------------------------------------------------------------------------------------
where $\beta$ is the inverse of the temperature (\ref{eq:beta}), 
$B_{\psi}$ is the boundary at infinity of a $\psi$ constant hypersurface $\Sigma_{\psi}$.
$\bar{N}_L$ is the lapse function, 
$\sqrt{\bar{\sigma}}$ is the area element of $B_{\psi}$, 
$\bar{k}$ is the trace of the extrinsic curvature of $B_{\psi}$
embedded in $\Sigma_{\psi}$, $\bar{u}$ is the outward pointing unit normal to $B_{\psi}$, 
$\bar{\Theta}_{ij}$ is the extrinsic curvature of $\Sigma_{\psi}$ embedded in the spacetime manifold,
and $\bar{\Theta}$ is the trace of the extrinsic curvature.

Here, we note that $B_{\tau}$ in (\ref{eq:def-H}) 
or $B_{\psi}$ in (\ref{eq:def-tension}) are the boundary at infinity of a $\tau$ or $\psi$ constant hypersurface
and do not include the Misner strings.
In the evaluation, 
we consider the Euclidean section outside of the outer horizon and there is no fixed point of the isometry
generated by $\partial_{\psi}$. 
However, there are the Misner strings along $\theta=0$ and $\theta = \pi$.
If one wish to reconstruct the classical action (\ref{eq:Euclid-action}) or (\ref{eq:action-on-KKmonopole})
in this Hamiltonian formalism along the direction $\partial_{\psi}$, 
the surface terms on the Misner string should be considered~\cite{Hawking:1998jf}.

%%%%%%%%%%%%%%%%%%%%%%%%%%%%%%%%%%%%%%%%%%%%%%%%%%%%%%%%%%%%%%%%%%%%%%%%%%%%%%%%%%%%%%%%%
\subsection{Using the flat background}
%%%%%%%%%%%%%%%%%%%%%%%%%%%%%%%%%%%%%%%%%%%%%%%%%%%%%%%%%%%%%%%%%%%%%%%%%%%%%%%%%%%%%%%%%%

Now, we consider the boundary adjusted flat background (\ref{eq:flatBG}).
Then, the Hamiltonian is calculated as
%--------------------------------------------------------------------------
\begin{eqnarray}
H = \frac{\pi}{2G} r_{\infty} \left( 2\rho_+ - \rho_- +\rho_0 \right),
\end{eqnarray}
%--------------------------------------------------------------------------
which is also interpreted as a relative quantity to the flat background.
The gravitational tension of the SqKK black hole along $\partial_{\psi}$ 
is calculated as
%----------------------------------------------
\begin{eqnarray}
\mathcal{T} = \frac{1}{4 G}\left(\rho_+ +\rho_- +2\rho_0\right).
\end{eqnarray}
%----------------------------------------------
With this form of the tension, we can write down expressions for the first law as
%----------------------------------------------
\begin{eqnarray}
dF &=& -SdT -Qd\Phi +\mathcal{T} dL, 
\label{eq:first-law-F} \\
dH &=& TdS - Qd\Phi +\mathcal{T} dL, \label{eq:first-law-H} \\ 
dM_{AD} &=& TdS + \Phi dQ +\mathcal{T} dL, 
\label{eq:first-law-AD} 
\end{eqnarray} 
%----------------------------------------
where $L:=2\pi r_{\infty}$ is the size of the $S^1$ fiber at infinity. 
The quantities $F$, $H$ and $M_{AD}$ are related to each other by the 
Legendre transformations as
%--------------------------------------------------------------------
\begin{eqnarray}
F = H -TS = M_{AD}-Q\Phi-TS.
\label{eq:Legendre-flat-AD}
\end{eqnarray}
%----------------------------------------------------------------------
Therefore, $F$, $H$ and $M_{AD}$ are thermodynamic potentials having 
$(T,\Phi,L)$, $(S,\Phi,L)$ and $(S,Q,L)$ as 
natural variables, respectively.
The transformations (\ref{eq:Legendre-flat-AD}) 
are summarized as follows:
\newcommand{\mapright}[1]{\smash{\mathop{\hbox to 1.5cm{\rightarrowfill}}\limits^{#1}}}
\newcommand{\mapleft}[1]{\smash{\mathop{\hbox to 1.5cm{\leftarrowfill}}\limits_{#1}}}
\newcommand{\mapdown}[1]{\Big\downarrow\llap{$\vcenter{\hbox{$\scriptstyle#1\,$}}$ }}
\newcommand{\mapup}[1]{\Big\uparrow\rlap{$\vcenter{\hbox{$\scriptstyle#1$}}$ }}
%-----------------------------------------------------------------------------------
\begin{eqnarray}
\begin{array}{ccccc}
F & \mapright{T\to S} & H & \mapright{\Phi \to Q}  & M_{AD}, 
\end{array} 
\end{eqnarray}
%-------------------------------------------------------------------------------------
where the long arrow means Legendre transformation between thermodynamic 
potentials and the small arrow on the long arrow represents conversion of 
variables. 
The first law (\ref{eq:first-law-F}) is quite natural because 
in the evaluation of the free energy $F$, we fixed the temperature, 
the electro-static potential and the 
size of the extra-dimension at the boundary, and hence, the free energy should be
associated with thermodynamic environment with fixed $(T, \Phi, L)$.

The Legendre transformations of $H$ or $M_{AD}$ with respect to $\mathcal{T}L$
does not give the Komar mass.
It implies that 
the gravitational tension is not a thermodynamic variable related to the 
Komar mass.
Instead, $\epsilon$ and $\Sigma$ defined as
%--------------------------------------------------------------
\begin{eqnarray}
\epsilon = L^2, \quad 
\Sigma = \frac{\mathcal{T}}{2L}= \frac{1}{16\pi G r_{\infty} } (\rho_++\rho_- 
+2\rho_0  ),
\end{eqnarray}
%------------------------------------------------------------
are a pair of thermodynamic variables related to the Komar mass.
Actually, these quantities satisfy
%---------------------------------------------------
\begin{eqnarray}
dM_K = TdS +\Phi dQ - \epsilon d\Sigma,
\label{eq:first-law-Komar}
\end{eqnarray}
%---------------------------------------------------
and 
%--------------------------------------------------------
\begin{eqnarray}
 F = M_K -TS - Q\Phi + \epsilon \Sigma.
 \label{eq:Legendre-Komar}
\end{eqnarray}
%-------------------------------------------------------------
Thus, the Komar mass can be interpreted as a thermodynamic potential with natural variables ($S, Q, \Sigma$).
Furthermore, from the Legendre transformations (\ref{eq:Legendre-flat-AD}) 
and (\ref{eq:Legendre-Komar}),
$\epsilon$ can be a thermodynamic variable of $F$, $H$ and $M_{AD}$.  
That is,
%---------------------------------------------------------------------------
\begin{eqnarray}
dF &=& -SdT-Qd\Phi +\Sigma d\epsilon, \label{eq:F-epsilon}\\
dH &=& TdS-Qd\Phi +\Sigma d\epsilon, \\
dM_{AD} &=& TdS+\Phi dQ +\Sigma d\epsilon.
\end{eqnarray}
%---------------------------------------------------------------------------
These expressions are consistent with the interpretation that $F$, $H$ or $M_{AD}$ is the thermodynamic potential 
with natural variables $(T, \Phi, L)$, $(S, \Phi, L)$ or $(S, Q, L)$, because the system with fixed $L$ 
is equivalent to that with fixed $\epsilon$ due to the relation $\epsilon \propto L^2$.

Using the quantities ($\epsilon, \Sigma$), the Legendre transform of the Hamiltonian gives
%---------------------------------------------------------------------------
\begin{eqnarray}
H-\epsilon \Sigma = \frac{3}{2}TS.
\label{eq:Legendre-H-W}
\end{eqnarray}
%-------------------------------------------------------------------------
The Legendre transformations (\ref{eq:Legendre-flat-AD}), 
(\ref{eq:Legendre-Komar}) and (\ref{eq:Legendre-H-W}) are summarized as follows:
%-----------------------------------------------------------------------
\[
\begin{array}{ccccc}
F & \mapright{T\to S} & H & \mapright{\Phi \to Q} & M_{AD} \\
 & & \mapdown{\epsilon \to \Sigma} & & \mapdown{\epsilon \to \Sigma} \\
& & W & \mapright{\Phi \to Q} & M_K,
\end{array} 
\]
%---------------------------------------------------------------------------
where we set $W = \frac{3}{2}TS$. 
Note that the Legendre transformation between $M_K$ and $W$ is nothing but the Smarr-type formula 
(\ref{eq:Smarr}).
The quantity $W$ can also be interpreted as a thermodynamic potential 
with natural variables ($S, \Phi, \Sigma$). 

%%%%%%%%%%%%%%%%%%%%%%%%%%%%%%%%%%%%%%%%%%%%%%%%%%%%%%%%%%%%%%%%%%%%%%%%%%%%%%%%%%%%%
\subsection{Using the KK monopole background}
%%%%%%%%%%%%%%%%%%%%%%%%%%%%%%%%%%%%%%%%%%%%%%%%%%%%%%%%%%%%%%%%%%%%%%%%%%%%%%%%%%%

Next, we evaluate the thermodynamic potentials and the gravitational tension 
by use of the boundary matched KK monopole background.
The Hamiltonian of the SqKK black hole relative to the KK monopole background is calculated as 
%-----------------------------------
\begin{eqnarray}
\tilde{H} = \frac{\pi}{2G} r_{\infty} \left(2\rho_+ -\rho_- +\rho_0 -\frac{r_{\infty}}{2} \right),
\end{eqnarray}
%-----------------------------------
which is related to the free energy on the KK monopole background as 
%---------------------------------------------------------------------
\begin{eqnarray}
\tilde{F} = \tilde{H} -TS.
\end{eqnarray}
%--------------------------------------------------------------------
This relation is the same as in the previous case using the flat background.

The AD mass (\ref{eq:ADmass-BH-flat}) is calculated using the flat background. 
One might think that the KK monopole is useful as a reference spacetime for the AD mass, 
but AD mass calculated by use of the boundary matched KK monopole spacetime diverges. 
Indeed, in the evaluation of the finite AD mass obtained in \cite{Cai:2006td}, 
the boundary of the KK monopole was not matched. 
However, its physical meaning is not clear because the disagreement of 
the period of the imaginary time between 
the SqKK black hole and the reference spacetime implies 
the disagreement in temperature, and thus the system is not 
in thermodynamic equilibrium. 
Actually, as discussed in \cite{Cai:2006td}, the AD mass evaluated 
on the boundary \lq\lq unmatched\rq\rq \ 
KK monopole 
does not seem to be any sensible thermodynamic quantity 
because it does not satisfy any consistent expression for the first law. 
Therefore, we do not consider the AD mass in the thermodynamic formulation with the KK monopole background.

The gravitational tension using the boundary matched KK monopole background is 
%-------------------
\begin{eqnarray}
\tilde{\mathcal{T}} &=& \frac{1}{4 G} \left( \rho_+ +\rho_- + 2\rho_0 -r_{\infty}\right),
\end{eqnarray}
%----------------------------
which contributes to the following expressions for the first law: 
%------------------------------------------------
\begin{eqnarray}
d\tilde{F} &=& -SdT -Qd\Phi + \tilde{\mathcal{T}} dL, 
\label{eq:F-first-KK} \\
d\tilde{H} &=& TdS -Qd\Phi + \tilde{\mathcal{T}} dL.
\end{eqnarray}
%-----------------------------------------------

As in the case of the flat background,
the Legendre transformation with respect to $\mathcal{T}L$ does not give the 
Komar mass, and $\mathcal{T}$ and $L$ are not related to the first law 
including the Komar mass. Instead, 
%--------------------------------------------------------------
\begin{eqnarray}
\epsilon = L^2, \quad \tilde{\Sigma} = \frac{\tilde{\mathcal{T}}}{2L}= \frac{1}{16\pi G r_{\infty} } (\rho_++\rho_- 
+2\rho_0  -r_{\infty})
\end{eqnarray}
%------------------------------------------------------------
are a couple of variables related to the Komar mass, because the following differential relation is satisfied:
%-----------------------------------------------------------
\begin{eqnarray}
dM_K = TdS +\Phi dQ - \epsilon d \tilde{\Sigma}.
\label{eq:first-law-Komar-GPS}
\end{eqnarray}
%---------------------------------------------------
Furthermore, the Legendre transformation 
between the free energy and the Komar mass is given as
%----------------------------------------------------------
\begin{eqnarray}
 \tilde{F} = M_K -TS - Q\Phi + \epsilon \tilde{\Sigma}.
\end{eqnarray}
%----------------------------------------------------------
One might think that the expression (\ref{eq:first-law-Komar-GPS}) 
conflicts with (\ref{eq:first-law-Komar})
because the Komar mass does not depend on 
the choice of the reference spacetime.
But, (\ref{eq:first-law-Komar-GPS}) and (\ref{eq:first-law-Komar}) are consistent,
because the difference between $\Sigma$ and $\tilde{\Sigma}$ 
is just a constant, i.e., $d\Sigma=d\tilde{\Sigma}$.
Again, $(\epsilon, \tilde{\Sigma})$ contribute to the differential 
relation satisfied by the free energy and the Hamiltonian as 
%------------------------------------------------
\begin{eqnarray} 
d\tilde{F} &=& -SdT -Qd\Phi + \tilde{\Sigma} d\epsilon, \\
d\tilde{H} &=& TdS -Qd\Phi + \tilde{\Sigma} d\epsilon.
\end{eqnarray}
%-----------------------------------------------
Furthermore, the Legendre transform of the Hamiltonian with respect to 
$\epsilon\tilde{\Sigma}$ gives $W$.
If we introduce a new quantity $\tilde{M}$ as the Legendre transform of 
$\tilde{H}$ with respect to $\Phi$:
%-------------------------------------------------------------
\begin{eqnarray}
\tilde{M} = \tilde{H} +Q\Phi,
\end{eqnarray}
%-----------------------------------------------------------------
it satisfies 
%-------------------------------------------------------------
\begin{eqnarray}
d\tilde{M} =  TdS +\Phi dQ + \tilde{\Sigma} d\epsilon.
\end{eqnarray}
%-----------------------------------------------------------------
Thus, $\tilde{M}$ can be considered as a thermodynamic potential 
having $(S, Q, \epsilon)$ as natural variables.
Therefore, $\tilde{M}$ corresponds to $M_{AD}$ of the flat background.

The Legendre transformations between thermodynamic potentials evaluated on the KK monopole background are
summarized as 
%------------------------------------------------------------
\[
\begin{array}{ccccc}
\tilde{F} & \mapright{T\to S} & \tilde{H} &  \mapright{ \Phi \to Q} & \tilde{M} \\
 & & \mapdown{\epsilon \to \tilde{\Sigma}}  &   & \mapdown{ \epsilon \to \tilde{\Sigma} } \\
  & & W & \mapright{\Phi \to Q} & M_K.
\end{array} 
\]
%---------------------------------------------------------------

In the formulation using the KK monopole background, we can take the 
$r_{\infty} \to \infty$ limit, 
under which the $\tau$ constant hypersurface becomes asymptotically 
Euclid space.
As is expected, in this limit, $\epsilon \tilde{\Sigma}$ and 
$\tilde{\mathcal{T}} L$ become zero, and $M_K$, $\tilde{F}$ and $\tilde{H}$ 
become those of five-dimensional Reissner-Nordstr\"om black hole 
evaluated on the five-dimensional Minkowski reference spacetime.
Thus, this formulation for the SqKK black hole includes the usual thermodynamic formulation for 
the five-dimensional Reissner-Nordstr\"om black hole as a limit.
In this sense, it is a generalized formulation of thermodynamics for five-dimensional electrically charged 
static black holes.

%%%%%%%%%%%%%%%%%%%%%%%%%%%%%%%%%%%%%%%%%%%%%%%%%%%%%%%%%%%%%%%%%%%%%%%%%%%%%%%%%%%%%%%%%%%%%%%%%%%%%%%%
\section{Thermodynamic potentials in various thermodynamic environments}
\label{section:potentials}
%%%%%%%%%%%%%%%%%%%%%%%%%%%%%%%%%%%%%%%%%%%%%%%%%%%%%%%%%%%%%%%%%%%%%%%%%%%%%%%%%%%%%%%%%%%%%%%%%%%%%%%%

In thermodynamics, fixing a set of variables corresponds to specifying 
a thermodynamic environment characterized by those variables.
In the previous section, we have obtained various thermodynamic quantities.
Each mass and free energy has a set of natural variables or control parameters 
and can be interpreted as a thermodynamic potential for the  environment 
characterized by the variables.

In this section, we obtain some free energies for different isothermal 
environments by use of the Legendre transformations, 
and discuss the state having the lowest value of the free energy, 
i.e., so-called the globally stable state in the sense of thermodynamics\footnote{
 Of cause, it is not proved that the known solutions are all solutions in the 
five-dimensional Einstein-Maxwell system. 
Therefore, the meaning of the global stability discussed 
here is restricted to the solutions treated in this paper.
}.
Generally, thermodynamical stability requires not only the global stability 
but also the local stability. 
However, we do not discuss the local stability of the SqKK black hole in 
this paper. This point will be mentioned in section \ref{section:summary}.

As was discussed in~\cite{Kurita:2007hu}, the thermodynamic environments 
with fixed $L$ and those with fixed $\epsilon$ are the same, 
but the thermodynamic environments with fixed $\mathcal{T}$ are different 
from that with fixed $\Sigma$.
Therefore, we treat $L$, $\mathcal{T}$ and $\Sigma$ as independent 
thermodynamic variables related to the compact extra dimension.

As for the thermodynamic variable with respect to the electric field, 
we consider only $\Phi$ in this section, 
because the quantity $Q\Phi$ is positive definite and 
the Legendre transformation by use of $Q\Phi$ does not change 
the relative order between the thermodynamic potentials. 
Therefore, the globally stable state in each environment with fixed $Q$ is 
the same as that in the environment with fixed $\Phi$.

%________________________________________________________________________________________
%~^^^^^^^^^^^^^^^^^^^^^^^^^^^^^^^^^^^^^^^^^^^^^^^^^^^^^^^^^^^^^^^^^^^^^^^^^^^^^^^^^^^^^^^
\subsection{Environment with fixed $(T,\Phi,L)$}
%________________________________________________________________________________________
%^^^^^^^^^^^^^^^^^^^^^^^^^^^^^^^^^^^^^^^^^^^^^^^^^^^^^^^^^^^^^^^^^^^^^^^^^^^^^^^^^^^^^^^^

As discussed in section \ref{sec:actions}, the equations (\ref{eq:first-law-F}) 
and (\ref{eq:F-first-KK}) show that the free energy $F$ or $\tilde{F}$ 
determines the suitable thermal state for the thermodynamic environment 
with fixed $(T,\Phi,L)$. 
Note that $F$, which is the free energy of the SqKK black hole relative to 
the flat background, is positive definite.
Hence the free energy of the flat background is the lowest,
which implies that it is the globally stable in this environment.

%~~~~~~~~~~~~~~~~~~~~~~~~~~~~~~~~~~~~~~~~~~~~~~~~~~~~~~~~~~~~~~~~~~~~~~~~~~~~~~~~~~~~~~~~~~
\subsection{Environment with fixed $(T,\Phi,\mathcal{T})$ }
%__________________________________________________________________________________________

The thermodynamic potential available for environment with fixed $(T,\Phi,\mathcal{T})$
is given by the Legendre transform of $F$ with respect to $L$, 
%----------------------------------------------------------------------------------------
\begin{eqnarray}
\Omega_{T\Phi\mathcal{T}} = F-\mathcal{T}L = -\frac{\pi}{2G} (\rho_- + \rho_0 )r_{\infty}.
\label{eq:Omega-TPhi-tension-flat}
\end{eqnarray}
%-----------------------------------------------------------------------------------------
This potential satisfies
%----------------------------------------------------------------------------------------
\begin{eqnarray}
d\Omega_{T\Phi\mathcal{T}} = -SdT-Qd\Phi - Ld\mathcal{T}.
\end{eqnarray}
%-----------------------------------------------------------------------------------------
Thus, $\Omega_{T\Phi\mathcal{T}}$ is the thermodynamic potential 
having $(T,\Phi,\mathcal{T})$ as natural variables. 
We note that $\Omega_{T\Phi\mathcal{T}}$ is always negative, which implies that 
the SqKK black hole may more preferable than the flat background in this environment.

The thermodynamic potential of the KK monopole relative to the flat background can be obtained 
by taking the KK monopole limit $\rho_{\pm}\to 0$ with $\rho_0>0$ as 
%---------------------------------------------------------------------------
\begin{eqnarray}
\hat{\Omega}_{T\Phi\mathcal{T}} = -\frac{\pi}{4G}r_{\infty}^2. 
\label{eq:Omega-TPhi-tension-KK-flat}
\end{eqnarray} 
%---------------------------------------------------------------------------
This is also negative definite.
Thus, the KK monopole is more preferable than the flat background. 
The potential of the SqKK black hole relative to the KK monopole is 
%-------------------------------------------------------------------------------
\begin{eqnarray}
\tilde{\Omega}_{T\Phi{\mathcal{T}}} = \tilde{F} -\tilde{\mathcal{T}} L 
 = \frac{\pi}{2G}r_{\infty} \left(-\rho_- - \rho_0 + \frac{r_{\infty}}{2} \right),
\label{eq:Omega-T-Phi-tension-BH-KKM}
\end{eqnarray}
%-------------------------------------------------------------------------------
which satisfies
%--------------------------------------------------------------------------------
\begin{eqnarray}
d\tilde{\Omega}_{T\Phi{\mathcal{T}}} = -SdT-Qd\Phi- Ld\tilde{\mathcal{T}}.
\end{eqnarray}
%----------------------------------------------------------------------------------
From (\ref{eq:Omega-TPhi-tension-flat}), 
(\ref{eq:Omega-TPhi-tension-KK-flat}) and 
(\ref{eq:Omega-T-Phi-tension-BH-KKM}), 
it is checked that the thermodynamic potentials are well-defined 
relative quantities; i.e., 
%---------------------------------------------------------------------------
\begin{eqnarray}
\Omega_{T\Phi\mathcal{T}} = \tilde{\Omega}_{T\Phi\mathcal{T}} +\hat{\Omega}_{T\Phi\mathcal{T}}.
\end{eqnarray}
%----------------------------------------------------------------------------
Therefore, as in the previous case, we can determine the most stable state among these solutions
as one that has the lowest value of the potential.
Now, $\tilde{\Omega}_{T\Phi{\mathcal{T}}}$ is positive definite, because
%--------------------------------------------------------------------------------
\begin{eqnarray}
\tilde{\Omega}_{T\Phi\mathcal{T}} = \frac{\pi}{2G} r_{\infty}\sqrt{\rho_- + \rho_0} 
(\sqrt{\rho_+ + \rho_0} - \sqrt{\rho_- + \rho_0})>0.
\end{eqnarray}
%--------------------------------------------------------------------------------------
Therefore, the KK monopole is the most stable among these solutions.

%~~~~~~~~~~~~~~~~~~~~~~~~~~~~~~~~~~~~~~~~~~~~~~~~~~~~~~~~~~~~~~~~~~~~~~~~~~~~~~~~~~
\subsection{Environment with fixed $(T, \Phi, \Sigma)$}
%____________________________________________________________________________________

Now, we consider thermodynamic environment controlled by $\Sigma$.
The thermodynamic potential relevant for this environment is 
%-------------------------------------------------------------------------
\begin{eqnarray}
\Omega_{T\Phi\Sigma} = F-\epsilon \Sigma = \frac{\pi}{4G} r_{\infty} ( \rho_+ -\rho_-),
\label{eq:thermodynamic-potential-BH-flat-T-Phi-Sigma}
\end{eqnarray}
%---------------------------------------------------------------------
which is positive definite and satisfies
%-------------------------------------------------------------------------
\begin{eqnarray}
d\Omega_{T\Phi\Sigma} = -SdT -Qd\Phi -\epsilon d\Sigma.
\end{eqnarray}
%-----------------------------------------------------------------------
The potential of the SqKK black hole relative to the KK monopole is
%------------------------------------------------------------------------
\begin{eqnarray}
\tilde{\Omega}_{T\Phi\Sigma} = \tilde{F} -\epsilon \tilde{\Sigma} = \frac{\pi}{4G} r_{\infty} (\rho_+ -\rho_-).
\end{eqnarray}
%-------------------------------------------------------------------------
It is seen that $\tilde{\Omega}_{T\Phi\Sigma}$ equals $\Omega_{T\Phi\Sigma}$, 
which implies that the potential of the KK monopole relative to the flat background equals zero. 
Actually, by taking the KK monopole limit in (\ref{eq:thermodynamic-potential-BH-flat-T-Phi-Sigma}), 
we obtain
%-----------------------------------------------------------------------------------
\begin{eqnarray}
\hat{\Omega}_{T\Phi\Sigma} = 0.
\end{eqnarray}
%------------------------------------------------------------------------------------------
Therefore, in this environment, the flat background and the KK monopole 
can be always in phase equilibrium.
From the viewpoint of statistical mechanics, it means that 
the contribution of these solutions to the partition function is the same.  
Both the flat background and the KK monopole are always the most stable.

Furthermore, for the extremal SqKK black hole, i.e., $\rho_+ = \rho_-$, 
where horizons are degenerate and temperature becomes zero,  
it is clear that $\Omega_{T\Phi\Sigma} = \tilde{\Omega}_{T\Phi\Sigma}=0$. 
Therefore, if the extremal SqKK black hole is locally stable, 
the flat background, the KK monopole and the extremal SqKK black hole can 
be in phase equilibrium in the environment with $T=0$.

%%%%%%%%%%%%%%%%%%%%%%%%%%%%%%%%%%%%%%%%%%%%%%%%%%%%%%%%%%%%%%%%%%%%%%%%%%%%%%%%%%%%%%%%%%%%%%%%%%%%%%%%
\section{Application to the black string}
\label{section:BS}
%%%%%%%%%%%%%%%%%%%%%%%%%%%%%%%%%%%%%%%%%%%%%%%%%%%%%%%%%%%%%%%%%%%%%%%%%%%%%%%%%%%%%%%%%%%%%%%%%%%%%%%%

Now, we investigate thermodynamical formulation for the charged black string 
using the two reference spacetimes: 
the flat background and the KK monopole. 
The metric of the black string is known as~\cite{Horowitz:2002ym}
%---------------------------------------------------------------------------------------------
\begin{eqnarray}
ds^2 = - V(\rho) d\tau^2 +\left( 1-\frac{\rho_+}{\rho} \right)^{-1} d\rho^2
 +\rho^2\left(1-\frac{\rho_-}{\rho} \right)
d\Omega^2 + \left(1-\frac{\rho_-}{\rho}\right)^{-1} dz^2,
\end{eqnarray}
%-----------------------------------------------------------------------------------------------
where $V(\rho)= (1-\frac{\rho_+}{\rho})(1-\frac{\rho_-}{\rho})$ and  $z$ 
has a range $0\le z\le L$. 
Then, $L$ is the size of the extra-dimension at infinity.
The gauge field is given by (\ref{eq:-Maxwell-gauge-field}).
The black string is characterized by three parameters $\rho_+, \rho_-$ and $L$.

Firstly, we obtain thermodynamic quantities of the black string and investigate the first law, 
and then calculate the thermodynamic potential of the black string relative to the most stable state 
in each environment.

The thermodynamical quantities:
the ADM mass, inverse of the temperature, entropy, electric charge and electrostatic potential 
of the black string are given by
%----------------------------------------------------------------
\begin{eqnarray}
M_{ADM} &=& 
 \frac{L}{4G} ( 2\rho_+ +\rho_- ), \\
\beta &=& T^{-1}= 4\pi \rho_+ \sqrt{\frac{\rho_+}{\rho_+-\rho_-}}, \\
S &=& \frac{A_H}{4G} = \frac{\pi L}{G}\rho_+^2\sqrt{1-\frac{\rho_-}{\rho_+} }, \\
Q &=& \pm \frac{\sqrt{3}L}{2G}\sqrt{\rho_+ \rho_- }, \\
\Phi &=& \pm \frac{\sqrt{3}}{2}\sqrt{\frac{\rho_- }{\rho_+}}.
\end{eqnarray}
%-------------------------------------------------------------
Unlike the SqKK black hole, the boundary topology 
of the flat background is the same as that of the black string. 
If the boundary of the black string is set at the surface $\rho=\rho_B$, the metric of the boundary matched 
flat background is
%-----------------------------------------------------------------------------------
\begin{eqnarray}
ds^2 = 
V(\rho_B) d\tau_E^2 +d\rho^2 +\rho^2 d\Omega_{S^2}^2+\left(1-\frac{\rho_-}{\rho_B}\right)^{-1}dz^2.
\end{eqnarray}
%------------------------------------------------------------------------------------
Here, the boundary of the flat background is set at 
%-----------------------------------------------------------------------
\begin{eqnarray}
\rho_{BR}=\rho_B\left(1-\frac{\rho_-}{\rho_B}\right)^{1/2},
\end{eqnarray}
%----------------------------------------------------------------
so as to match the circumference radius of the $S^2$. 
Then, the classical action relative to the flat background is 
%---------------------------------------------------------------------
\begin{eqnarray}
I_E = \frac{L}{4G} \beta (\rho_+ -\rho_- ).
\end{eqnarray}
%-------------------------------------------------------------------
The free energy is obtained as
%------------------------------------------------------------------------
\begin{eqnarray}
F = \frac{I_E}{\beta} =  \frac{L}{4G} (\rho_+ - \rho_- ).
\end{eqnarray}
%-------------------------------------------------------------------------
With this reference spacetime, the Hamiltonian and gravitational tension of 
the black string can be calculated as 
%-------------------------------------------------------------------------
\begin{eqnarray}
H = \frac{L}{2G}(\rho_+ -\rho_- ), \quad 
\mathcal{T} = \frac{1}{4G}(\rho_+-\rho_- ).
\end{eqnarray}
%---------------------------------------------------------------------------
As is well-known, the black string has the gravitational tension as a thermodynamic quantity 
and the following relations are satisfied: 
%-----------------------------------------------------------------
\begin{eqnarray}
d F &=& -SdT - Qd\Phi +\mathcal{T} dL, \\
d H &=& TdS -Q d \Phi + \mathcal{T} dL,  \\
d M_{ADM} &=& TdS + \Phi dQ + \mathcal{T} dL.
\end{eqnarray}
%-------------------------------------------------------------------
These potentials are related each other via the Legendre transformations as
%---------------------------------------------------------------------
\begin{eqnarray}
F = H -TS = M_{ADM} -Q\Phi- TS.
\label{eq:Legendre-string-flat}
\end{eqnarray}
%---------------------------------------------------------------
The five-dimensional Komar mass of the black string is calculated as
%--------------------------------------------------------------
\begin{eqnarray}
M_K = - \frac{3}{32\pi G}  \int dS_{\mu\nu} \nabla^{\mu}\xi^{\nu} = \frac{3}{8 G} L(\rho_+ + \rho_-).
\end{eqnarray}
%-------------------------------------------------------------------
As is the case of the SqKK black hole, the quantities 
%-----------------------------------------------------------------------
\begin{eqnarray}
\epsilon = L^2, \quad\mbox{and}\quad \Sigma = \frac{\mathcal{T} }{2 L} = \frac{1}{8GL} (\rho_+ -\rho_- )
\end{eqnarray}
%-----------------------------------------------------------------------
are relevant for the Komar mass, in the sense that the Komar mass satisfies
%-----------------------------------------------------------------------
\begin{eqnarray}
dM_K = TdS +\Phi dQ -\epsilon d\Sigma.
\end{eqnarray}
%-----------------------------------------------------------------------
Furthermore, the relation $M_{ADM} = M_K + \epsilon \Sigma$ gives
another expression for the first law for the ADM mass: 
%-----------------------------------------------------------------------
\begin{eqnarray}
dM_{ADM } = TdS +\Phi dQ +\Sigma d\epsilon.
\end{eqnarray}
%------------------------------------------------------------------
Then, the Legendre transformations (\ref{eq:Legendre-string-flat}) leads
%---------------------------------------------------------------
\begin{eqnarray}
dF &=& -SdT -Qd\Phi +\Sigma d\epsilon, \\
dH &=& TdS -Qd\Phi +\Sigma d\epsilon.
\end{eqnarray}
%-----------------------------------------------------------------
Again, the Smarr-type relation is obtained as
%-----------------------------------------------------------------
\begin{eqnarray}
M_K -Q\Phi = \frac{3}{2}TS =: W.
\end{eqnarray}
%-----------------------------------------------------------------
The relations between thermodynamic potentials are summarized as
%-----------------------------------------------------------------------
\[
\begin{array}{ccccc}
F & \mapright{T\to S} & H & \mapright{\Phi \to Q} & M_{ADM} \\
 & & \mapdown{\epsilon \to \Sigma} & & \mapdown{\epsilon \to \Sigma} \\
& & W & \mapright{\Phi \to Q} & M_K.
\end{array} 
\]
%---------------------------------------------------------------------------
This is the same as the case of the SqKK black hole, where the ADM mass 
of the string corresponds to the AD mass of the SqKK black hole.

The KK monopole background can be useful as a reference spacetime of the 
black string, though the boundary topology of the black string is not 
the same as that of the KK monopole. 
We set the boundary of the black string at the surface $\rho=\rho_B$.
The metric of the KK monopole can be rewritten as 
%------------------------------------------------------------------------
\begin{eqnarray}
ds_E^2 = V(\rho_B) d\tau_E^2+B_R(\rho)d\rho^2 + \rho^2 B_R(\rho)d\Omega_{S^2}^2 + 
\frac{\rho^2}{B_R (\rho)} (d\psi + \cos\theta d\phi)^2,
\label{metric:KK-string}
\end{eqnarray}
%-----------------------------------------------------------------------------------
where $B_R(\rho) = 1+\frac{\rho_{0R}}{\rho}$. The boundary of the KK monopole is set at surface 
$\rho=\rho_{BR}$. 
In order to match the period of the imaginary time and the size of $S^1$ at the boundary, 
we set parameters $\rho_{BR}$ and $\rho_{0R}$ as 
%------------------------------------------------------------------------------
\begin{eqnarray}
\rho_{BR} &=& \rho_B \left(1-\frac{\rho_-}{\rho_B} - \frac{L}{4\pi\rho_B}\right)^{1/2}, \\
\rho_{0R} &=& \frac{L}{4\pi} \frac{\rho_B}{\rho_{BR}}. 
\end{eqnarray}
%------------------------------------------------------------------------------
Then, the finite classical action is obtained as 
%--------------------------------------------------------------------------
\begin{eqnarray}
I_E = \frac{L}{4G}\beta \left(\rho_+ - \rho_- - \frac{L}{4\pi} \right).
\end{eqnarray}
%-----------------------------------------------------------------------------
The free energy of the black string relative to the KK monopole is 
%------------------------------------------------------------------------------
\begin{eqnarray}
\tilde{F} = \frac{L}{4G} \left(\rho_+ - \rho_- - \frac{L}{4\pi} \right).
\end{eqnarray}
%----------------------------------------------------------------------------
The Hamiltonian and the gravitational tension can also be calculated 
by use of the KK monopole (\ref{metric:KK-string}) as reference spacetime as 
%-----------------------------------------------------------------------
\begin{eqnarray}
\tilde{H} = \frac{L}{2G} \left( \rho_+ - \rho_- -\frac{L}{8\pi} \right),\quad 
\tilde{\mathcal{T}} = \frac{1}{4G} \left(\rho_+ -\rho_- -\frac{L}{2\pi} \right).
\end{eqnarray}
%------------------------------------------------------------------
Using these quantities, the free energy and Hamiltonian satisfy 
%-----------------------------------------------------------------------
\begin{eqnarray}
d\tilde{F} & = & -SdT -Qd\Phi + \tilde{\mathcal{T}} dL, \\
d\tilde{H} & = & TdS -Qd\Phi + \tilde{\mathcal{T}} dL.
\end{eqnarray}
%----------------------------------------------------------------------
The expression of the first law by the Komar mass can be written as
%-------------------------------------------------------------
\begin{eqnarray} 
dM_K = TdS +\Phi dQ - \epsilon d\tilde{\Sigma},
\end{eqnarray}
%--------------------------------------------------------------------
where
%----------------------------------------------------------------
\begin{eqnarray}
\epsilon:= L^2, \quad 
\tilde{\Sigma} := \frac{\tilde{\mathcal{T}}}{2L}= \frac{1}{8GL}\left(
\rho_+ - \rho_- -\frac{L}{2\pi}
\right).
\end{eqnarray}
%--------------------------------------------------------------------------
The thermodynamic potentials are related by the following Legendre transformations:
%--------------------------------------------------------------
\begin{eqnarray}
M_K-Q\Phi = W 
 = \tilde{H} - \epsilon \tilde{\Sigma} 
= \tilde{F}+TS - \epsilon \tilde{\Sigma},
\end{eqnarray}
%-----------------------------------------------------------------
and thus, $\tilde{\Sigma}$ can contribute to 
the expression of the first law for $\tilde{F}$ and $\tilde{H}$ as 
%-----------------------------------------------------------------------------
\begin{eqnarray} 
d \tilde{F} &=& -SdT -Qd\Phi + \tilde{\Sigma} d\epsilon,  \\
d \tilde{H} &=& TdS -Qd\Phi +\tilde{\Sigma} d\epsilon.
\end{eqnarray}
%--------------------------------------------------------------------
If we introduce a new quantity $\tilde{M}$ as the Legendre transform 
of $\tilde{H}$ with respect to $\Phi$: 
%-------------------------------------------------------------
\begin{eqnarray}
\tilde{M} = \tilde{H} +Q\Phi = \frac{L}{4G}\left(
2\rho_+ + \rho_- - \frac{L}{4\pi}
\right),
\end{eqnarray}
%-----------------------------------------------------------------
it satisfies 
%-------------------------------------------------------------
\begin{eqnarray}
d\tilde{M} =  TdS +\Phi dQ + \tilde{\Sigma} d\epsilon.
\end{eqnarray}
%-----------------------------------------------------------------
Thus, $\tilde{M}$ can be considered as a thermodynamic potential having 
$(S, Q, \epsilon)$ as natural variables,
and is a counterpart of the AD mass with respect to the KK monopole background.

Therefore, we can write the following map for the black string with the KK monopole background:
%------------------------------------------------------------
\[
\begin{array}{ccccc}
\tilde{F} & \mapright{T\to S} & \tilde{H} &  \mapright{ \Phi \to Q} & \tilde{M} \\
 & & \mapdown{\epsilon \to \tilde{\Sigma}}  &   & \mapdown{ \epsilon \to \tilde{\Sigma} } \\
  & & W & \mapright{\Phi \to Q} & M_K.
\end{array} 
\]
%----------------------------------------------------------

Now, it is found that we can formulate the thermodynamics of 
the black string in a quite similar way 
to that of the SqKK black hole, where the thermodynamic environments are 
characterized by 
the same sets of variables as those in the case of the SqKK black hole. 
In the remaining part of this section, 
we calculate the thermodynamic potential of the black string 
in two thermodynamic environment with fixed 
($T,\Phi, \mathcal{T}$) and ($T, \Phi, \Sigma$), 
taking the most stable state in each environment
as a reference background.

When the set ($T,\Phi, \mathcal{T}$) is fixed, the most stable state is the KK monopole.
The potential of the black string relative to the KK monopole is given as
%------------------------------------------------------------------
\begin{eqnarray}
\tilde{\Omega}_{T\Phi\mathcal{T}} = \tilde{F} - \tilde{\mathcal{T}} L = \frac{L^2}{16\pi G}.
\label{eq:potential-string-Phi-mathcal{T}}
\end{eqnarray}
%------------------------------------------------------------------
This is positive definite.
It is found that the potential (\ref{eq:potential-string-Phi-mathcal{T}}) 
depends on neither $\rho_+$ nor $\rho_-$.
It implies that the potential relative to the flat background equals zero.
Actually, it is calculated as 
%---------------------------------------------------------------------
\begin{eqnarray}
\Omega_{T\Phi\mathcal{T}}:=F-\mathcal{T} L =0,
\end{eqnarray}
%---------------------------------------------------------------------
which always vanishes.
Thus, if the black string is locally stable in the sense of the thermodynamics, 
then the flat spacetime and the black string are in phase equilibrium.

This result holds generally for solutions without Misner string and bolt 
singularity with respect to isometry related to the gravitational tension.
The reason is as follows: 
The gravitational tension can be obtained as a Hamiltonian along isometry
in a compact spatial direction.
In such a Hamiltonian formulation, the classical Euclidean action can be 
reconstructed from the gravitational tension (Hamiltonian) 
and contribution from 
Misner strings 
and bolt singularities as \cite{Hawking:1998jf}
%-------------------------------------------
\begin{eqnarray}
I_E = \beta(L\mathcal{T} + H_{MS}) - \frac{1}{4G}(A_{bolts} + A_{MS}),
\end{eqnarray}
%----------------------------------------------
where $H_{MS}$ is Hamiltonian surface term on the Misner strings, $A_{bolts}$ and $A_{MS}$ are
respectively total area of the bolts and the Misner strings in the spacetime.
In the present case, there is no Misner string and bolt singularity in the spacetime 
as well as in the reference spacetime. Therefore, 
%-----------------------------------------------------------------
\begin{eqnarray}
I_E = \beta L \mathcal{T}, 
\end{eqnarray}
%-------------------------------------------------------------------
which means that the Legendre transform of $F(=I_E/\beta$) with respect to $L$ , which is 
thermodynamic potential for isothermal environment with fixed gravitational tension, vanishes.

When $\Sigma$ is a variable of the system, the KK monopole background is as preferable as
the flat background and either reference spacetime gives the same result.
Here, we consider the flat background.
For example, the potential relevant for the environment with fixed ($T, \Phi, \Sigma$) is 
%---------------------------------------------------------
\begin{eqnarray}
{\Omega}_{T\Phi\Sigma} = {F} -\epsilon {\Sigma} = \frac{L}{8G} (\rho_+-\rho_-),
\end{eqnarray}
%----------------------------------------------------
which is positive definite. 
As in the case of the SqKK black hole, for an extremal black string, i.e., 
$\rho_+ = \rho_-$, the potential vanishes.

As shown in this section, the KK monopole can be a background spacetime for the
black string as well as the flat background.
Especially, for thermodynamic environment with fixed ($T, \Phi, \mathcal{T}$), 
the KK monopole should be considered as a reference spacetime 
because the potential relative to the flat background is always negative.

%%%%%%%%%%%%%%%%%%%%%%%%%%%%%%%%%%%%%%%%%%%%%%%%%%%%%%%%%%%%%%%%%%%%%%%%%%%%%%%%%%%%%%%%%%%%%%%%%%%%%%%%
\section{Summary and discussion}
\label{section:summary}
%%%%%%%%%%%%%%%%%%%%%%%%%%%%%%%%%%%%%%%%%%%%%%%%%%%%%%%%%%%%%%%%%%%%%%%%%%%%%%%%%%%%%%%%%%%%%%%%%%%%%%%%

The formulation of thermodynamics of the charged squashed Kaluza-Klein 
(SqKK) black hole and the charged black string has been investigated by 
the background subtraction scheme with two reference spacetimes. 
The natural reference spacetime for the SqKK black hole is the KK monopole 
because the boundaries of these spacetimes have the same topology. 
With this reference spacetime, we can take the spherically symmetric limit 
and reproduce the usual thermodynamic formulation for the five-dimensional 
Reissner-Nordstr\"om black hole in the limit. 
It has been shown for the SqKK black hole that 
the counterterm method is equivalent to the background subtraction method 
with the flat background,
in which the boundary topology is different from that of the SqKK black hole. 
As in the case of the Taub-bolt instanton,  
the counterterm method prefers the gravitational action without perfect matching.

We have seen that the free energy, the Hamiltonian, the AD mass and the Komar 
mass are related by the Legendre transformation as thermodynamic potentials.
By calculating thermodynamic potentials for a given thermodynamic environment,
we can determine the most stable state by these quantities. 
In the isothermal environments with fixed size of the extra 
dimension $L$, the flat background is the most stable. 
On the other hand, in those with fixed gravitational tension $\mathcal{T}$, 
the KK monopole is the most stable. 
Furthermore, when we fix $\Sigma$ which is conjugate to $\epsilon =L^2$, 
the flat background and the KK monopole can be in phase equilibrium and 
both are the most stable.

We have constructed a thermodynamic formulation for the five-dimensional
charged black string concretely, 
which shows that the thermodynamic variables are the 
same as that of the SqKK black hole.
It has been shown that the KK monopole is useful as a reference spacetime 
for the charged black string. 
In particular, the KK monopole should be considered as a reference spacetime
in isothermal environments with fixed gravitational tension, because
the KK monopole is the most stable.

We are considering several future works related to the present paper.
The thermodynamic formulation will be able to be generalized for 
the rotating charged SqKK black hole, which was recently found in 
\cite{Nakagawa:2008rm}. 
This would be a combination of the work in \cite{Wang:2006nw} and the present work.
Further, it is interesting to investigate thermodynamic stability of the SqKK black hole 
and the black string using thermodynamic formulation given in this paper, 
because there seems to be phase transition between the black string and 
the SqKK black hole and between the SqKK black hole and the flat background 
in some environments.
However, when we discuss phase transition, 
the property of the thermodynamical local stability must be clarified. 
If the entropy of a given system is additive, then the parameter region of 
local stability can be shown by examining the Hessian matrix of the entropy, 
but it is not trivial that black hole entropy is additive or not 
(see, for example, \cite{Arcioni:2004ww}).
Therefore, other methods to investigate the local stability should be applied to this problem.
We will report thermodynamic stability of these solutions in a future publication.

%%%%%%%%%%%%%%%%%%%%%%%%%%%%%%%%%%%%%%%%%%%%%%%%%%%%%%%%%%%%%%%%%%%%%%%%%%%%%%%%%%%%%%%%%%%%%%%%%%%%%%%%
\begin{acknowledgments}
YK is supported by the 21st Century COE "Constitution of wide-angle mathematical basis focused on knots" 
from the Ministry of Education, Culture,
Sports, Science and Technology (MEXT) of Japan.
This work is supported by the Grant-in-Aid for Scientific Research Fund of the Ministry of Education, Science and
Culture of Japan No. 19540305.
\end{acknowledgments} 
%%%%%%%%%%%%%%%%%%%%%%%%%%%%%%%%%%%%%%%%%%%%%%%%%%%%%%%%%%%%%%%%%%%%%%%%%%%%%%%%%%%%%%%%%%%%%%%%%%%%%%%%%

%%%%%%%%%%%%%%%%%%%%%%%%%%%%%%%%%%%%%%%%%%%%%%%%%%%%%%%%%%%%%%%%%%%%%%%%%%%%%%%%%%%%%%%%%%%%%%%%%

\end{document}